# From vanishing pairwise entanglement to global separability: Entanglement structure and measures in the *W* subspace

[1] Reza Hamzehofi

[1]Department of Physics, Faculty of Science, Shahid Chamran University of Ahvaz, Ahvaz, Iran
E-mail: rezahamzehofi@gmail.com

The *W* subspace is defined as the subspace of the $n$-qubit Hilbert space spanned by states containing at most one excitation. In this work, a separability criterion is established for pure and mixed states supported in this subspace: if all reduced two-qubit subsystems are separable, then the entire $n$-qubit state is necessarily separable. This result motivates the identification of the sum of two-tangles as an entanglement measure for pure states supported in the *W* subspace. Furthermore, it is shown that the commonly used $\pi$-tangle becomes ineffective for the $n$-qubit *W* state as the system size increases, and vanishes in the large-$n$ limit. To address this limitation, the sum of $\pi$-tangles is introduced, which, like the sum of two-tangles, successfully quantifiesthe entanglement of the $n$-qubit *W* state in the large-$n$ limit. In addition, a new condition for entanglement measures is introduced, which facilitates the formulation of a well-behaved and physically meaningful entanglement measure.

## I. Introduction

Genuine entanglement is a central resource in quantum information theory, enabling protocols such as quantum teleportation, quantum communication, and distributed quantum computation [1-4]. Among genuinely entangled states, the *W*-class occupies a special position because of its robustness: unlike GHZ-class states, the entanglement present in *W*-class states survives the loss of any subsystem and is manifested in the persistent bipartite correlations shared across all partitions. This structural property makes *W*-type entanglement especially valuable for realistic quantum networks, noisy communication channels, and physical architectures where particle loss or decoherence is unavoidable [5-7].

In a previous study with my collaborators, we showed that the pairwise entanglement of a *W* state, although decreasing with the number of qubits, never fully vanishes [8]. However, if a pure *W* state undergoes a physical evolution and becomes a mixed state, the pairwise entanglement may indeed disappear. For instance, when one of the qubits accelerates uniformly, at high accelerations the pairwise entanglement between that qubit and the others can vanish entirely [8]. This observation raises an important question: if a pure *W* state undergoes a physical evolution and becomes a mixed state, and all pairwise entanglements vanish, can one conclude that the total entanglement of the system is lost? While the answer might seem obvious at first glance, it is not trivial. In certain states, the pairwise entanglement can vanish while the system still retains genuine multipartite entanglement. For example, the GHZ-class states have this feature. Therefore, for the *W*-class states, a rigorous mathematical proof is required. One of the objectives of this study is to provide such a rigorous proof for the *W*-class states. To this end, The *W* subspace is defined as the subspace of the $n$-qubit Hilbert space spanned by states containing at most one excitation. We then investigate the relation between pairwise entanglement and global separability, showing that the vanishing of all pairwise entanglements implies complete separability for any state supported in the *W* subspace. Since all *W*-class states belong to this subspace, this result provides an affirmative answer to the aforementioned question.

Another objective of this study is to investigate whether, based on the above theorem, an entanglement measure can be defined for pure states supported in the *W* subspace as a function of pairwise entanglement. This question is particularly interesting because the pairwise entanglements are significantly easier to calculate than the entanglement of the entire multipartite system. Accordingly, the structure of this paper is as follows. In Section II, we introduce the main entanglement measures employed in this study. Section III presents a separability criterion established for pure and mixed states supported in the *W* subspace. In Section IV, using the result from Section III, an entanglement measure tailored to pure states supported in the *W* subspace is proposed. Finally, Section V summarizes the conclusions.

## II. Entanglement measures

For a two-qubit state $\rho_{AB}$, the concurrence $C_{AB}$ is defined as follows [9]:

$$C_{AB} = \max\{0, \lambda_1 - \lambda_2 - \lambda_3 - \lambda_4\}, \qquad (1)$$

where $\lambda_1 \geq \lambda_2 \geq \lambda_3 \geq \lambda_4$ are the square roots of the eigenvalues of the matrix $R = \rho_{AB}\left(\sigma_y \otimes \sigma_y\right)\rho^*_{AB}\left(\sigma_y \otimes \sigma_y\right)$. Here, $\rho^*_{AB}$ is the complex conjugate of $\rho_{AB}$ in the computational basis and $\sigma_y$ is the Pauli y-matrix.

To quantify the entanglement of a three-qubit pure state, the three-tangle $\tau_3$ is defined in terms of the concurrences as follows [10]:

$$\tau_3 = \tau_{A(BC)} - \tau_{AB} - \tau_{AC}, \qquad (2)$$

where the one-tangle $\tau_{A(BC)}$ denotes the squared concurrence between subsystem *A* and the composite subsystem *BC*, and the two-tangles $\tau_{AB}$ and $\tau_{AC}$ are the squared pairwise concurrences. Moreover, for a three-qubit pure state, we have: $\tau_{A(BC)} = 4\det\left(\rho_A\right)$. Another notable feature of *W*-class states is that, in contrast to GHZ-class states, they have vanishing three-tangle [10]. In other words, this measure cannot be used to quantify the entanglement of the *W*-class states.

If we rewrite Eq. (2) in terms of negativity instead of concurrence, the π-tangle is obtained. This measure is given by [11]:

$$\pi_3 = N^2_{A(BC)} - N^2_{AB} - N^2_{AC}, \qquad (3)$$

where $N_{A(BC)}$ is the negativity between qubit *A* and the composite subsystem *BC*. For pure states we have $N_{A(BC)} = C_{A(BC)}$. Moreover for a two-qubit state $\rho_{\alpha\beta}$, the negativity $N_{\alpha\beta}$ is defined as follows [12]:

$$N_{\alpha\beta} = 2\sum_i |\lambda_i| \quad \left(\lambda_i < 0\right), \qquad (4)$$

where $\lambda_i$ are the negative eigenvalues of the partial transpose matrix $\rho^{T_j}_{AB}$ with respect to the subsystem $j \in \{A, B\}$. Unlike the three-tangle, which vanishes for *W*-class states, the π-tangle takes a non-zero value [11].

## III. A separability criterion for states supported in the *W* subspace

For a three-qubit system, there are only two types of genuine entangled states: the GHZ-class states and the *W*-class state [13]. A symmetric *W*-class state is defined as follows [14]:

$$|W\rangle=\frac{1}{\sqrt{|a|^2+3}}\left(a|000\rangle+|001\rangle+|010\rangle+|100\rangle\right). \quad (5)$$

For $a=0$, this state transforms into the $W$ state. Additionally, the asymmetric $W$ state can be considered in the following form:

$$|W\rangle=k_1|001\rangle+k_2|010\rangle+k_3|100\rangle, \quad (6)$$

where $\sum_i |k_i|^2=1$.

Now, we define the $W$ subspace in an $n$-qubit Hilbert space $\mathcal{H}=\left(\mathbb{C}^2\right)^{\otimes n}$ as the subspace spanned by the computational basis states with at most one excitation, namely

$$\mathcal{H}_W=\text{span}\left\{|0\rangle^{\otimes n},|0\rangle^{\otimes n-1}|1\rangle,|0\rangle^{\otimes n-2}|1\rangle|0\rangle,\ldots,|1\rangle|0\rangle^{\otimes n-1}\right\}. \quad (7)$$

It is called the $W$ subspace because $W$-class states are supported in this subspace.

**Theorem 1:** Consider state $\rho_W$ supported in the $W$ subspace. If, the sum of two-tangles vanishes, then the state is completely separable.

**Proof:** Here, the theorem is first proved for the three-qubit case and then generalized to $n$-qubit states. In general, the density matrix of a three-qubit state supported in the $W$ subspace can be written as follows:

$$\rho_W=\begin{pmatrix} \alpha & x_1 & x_2 & x_3 \\ x_1^* & \beta_1 & g & h \\ x_2^* & g^* & \beta_2 & t \\ x_3^* & h^* & t^* & \beta_3 \end{pmatrix}, \quad (8)$$

where $\alpha+\sum_i \beta_i=1$. The density matrix is written in terms of the below basis

$$\{|000\rangle,|001\rangle,|010\rangle,|100\rangle\}. \quad (9)$$

Moreover, the reduced density matrices $\rho_W^{AB}$, $\rho_W^{AC}$, and $\rho_W^{BC}$ are listed below:

$$\rho_W^{AB}=\begin{pmatrix} \alpha+\beta_1 & x_2 & x_3 & 0 \\ x_2^* & \beta_2 & t & 0 \\ x_3^* & t^* & \beta_3 & 0 \\ 0 & 0 & 0 & 0 \end{pmatrix},$$

$$\rho_W^{AC}=\begin{pmatrix} \alpha+\beta_2 & x_1 & x_3 & 0 \\ x_1^* & \beta_1 & h & 0 \\ x_3^* & h^* & \beta_3 & 0 \\ 0 & 0 & 0 & 0 \end{pmatrix}, \quad (10)$$

$$\rho_W^{BC}=\begin{pmatrix} \alpha+\beta_3 & x_1 & x_2 & 0 \\ x_1^* & \beta_1 & g & 0 \\ x_2^* & g^* & \beta_2 & 0 \\ 0 & 0 & 0 & 0 \end{pmatrix}.$$

The three matrices above are written in the two-qubit Hilbert space in terms of the below basis

$$\{|00\rangle,|01\rangle,|10\rangle,|11\rangle\}. \quad (11)$$

Using Eqs. (1) and (10), the sum of two-tangles is obtained as follows:

$$C_{AB}^2+C_{AC}^2+C_{BC}^2=4\left(|t|^2+|h|^2+|g|^2\right). \quad (12)$$

It follows from the above relation that if the sum two-tangles vanishes, then we have:

$$t=h=g=0. \quad (13)$$

Now, condition (13) is applied to the density matrix in Eq. (8) to obtain the following matrix:

$$\rho_W'=\begin{pmatrix} \alpha & x_1 & x_2 & x_3 \\ x_1^* & \beta_1 & 0 & 0 \\ x_2^* & 0 & \beta_2 & 0 \\ x_3^* & 0 & 0 & \beta_3 \end{pmatrix}. \quad (14)$$

Next, it is necessary to show that the above matrix represents a separable mixed state. To do this, it is sufficient to write it in the following form [15]

$$\rho_{\text{ens}} = \sum_i p_i |\psi_i\rangle\langle\psi_i|, \tag{15}$$

where $\sum_i p_i = 1$.

Given the basis of the Hilbert space in Eq. (9), the vectors $|\psi_i\rangle$ are expected to have the following form:

$$\begin{aligned} |\psi_0\rangle &= |000\rangle, \\ |\psi_1\rangle &= \sqrt{1-|\upsilon_1|^2}\,|000\rangle + \upsilon_1|001\rangle, \\ |\psi_2\rangle &= \sqrt{1-|\upsilon_2|^2}\,|000\rangle + \upsilon_2|010\rangle, \\ |\psi_3\rangle &= \sqrt{1-|\upsilon_3|^2}\,|000\rangle + \upsilon_3|001\rangle. \end{aligned} \tag{16}$$

Now, the coefficients $p_i$ need to be determined. Using Eqs. (15) and (16), and comparing the result with the density matrix in Eq. (14), we obtain:

$$p_0 + \sum_{i=1}^{3} p_i |\upsilon_i|^2 = \alpha, \tag{17}$$

$$p_i |\upsilon_i|^2 = \beta_i \qquad i \in \{1,2,3\}, \tag{18}$$

$$p_i \upsilon_i \sqrt{1-|\upsilon_i|^2} = x_i \qquad i \in \{1,2,3\}. \tag{19}$$

From Eqs. (18) and (19), one can readily derive the following expression:

$$p_i = \beta_i + \frac{|x_i|^2}{\beta_i} \qquad i \in \{1,2,3\}. \tag{20}$$

Using the obtained relations, the vectors $|\psi_1\rangle$, $|\psi_2\rangle$, and $|\psi_3\rangle$ can be rewritten as follows:

$$\begin{aligned} |\psi_1\rangle &= \frac{1}{\sqrt{p_1}}\left(\frac{x_1}{\sqrt{\beta_1}}|000\rangle + \sqrt{\beta_1}\,|001\rangle\right), \\ |\psi_2\rangle &= \frac{1}{\sqrt{p_2}}\left(\frac{x_2}{\sqrt{\beta_2}}|000\rangle + \sqrt{\beta_2}\,|010\rangle\right), \\ |\psi_3\rangle &= \frac{1}{\sqrt{p_3}}\left(\frac{x_3}{\sqrt{\beta_3}}|000\rangle + \sqrt{\beta_3}\,|001\rangle\right). \end{aligned} \tag{21}$$

Now, using Eqs. (20) and (21), the density matrix in Eq. (15) takes the following form:

$$\rho_{ens} = \begin{pmatrix} \rho_{ens}^{1,1} & x_1 & x_2 & x_3 \\ x_1^* & \beta_1 & 0 & 0 \\ x_1^* & 0 & \beta_2 & 0 \\ x_1^* & 0 & 0 & \beta_3 \end{pmatrix}, \tag{22}$$

where $\rho_{ens}^{1,1} = p_0 + \frac{|x_1|^2}{\beta_1} + \frac{|x_2|^2}{\beta_2} + \frac{|x_3|^2}{\beta_3}$. By comparing the density matrices in Eqs. (14) and (22), it is found that, for the density matrix $\rho_{ens}$ to coincide with the density matrix in Eq. (14), the following condition must be satisfied:

$$\frac{|x_1|^2}{\beta_1} + \frac{|x_2|^2}{\beta_2} + \frac{|x_3|^2}{\beta_3} \le \alpha. \tag{23}$$

This inequality also ensures that $p_0$ is nonnegative, since the nonnegativity of this variable is necessary for expressing $\rho_{ens}$. Therefore, the variable $p_0$ can be computed as follows:

$$p_0 = \alpha - \frac{|x_1|^2}{\beta_1} - \frac{|x_2|^2}{\beta_2} - \frac{|x_3|^2}{\beta_3}. \tag{24}$$

To prove inequality (23), it should be noted that a density matrix is positive semidefinite. Therefore, all principal minors of the density matrix are nonnegative; in particular, its determinant is nonnegative [16]. It then follows that:

$$Det(\rho''_W) = \alpha\beta_1\beta_2\beta_3 - |x_1|^2 \beta_2\beta_3 - |x_2|^2 \beta_1\beta_3 - |x_3|^2 \beta_1\beta_2 \geq 0. \quad (25)$$

Since $\beta_i$ are nonnegative and real, and assuming that none of them is zero, dividing the entire expression by $\beta_1\beta_2\beta_3$ yields inequality (23).

Based on the Sylvester's criterion, if one or more $\beta_i$ is zero, the corresponding off-diagonal must be zero as well [17]: for example if $\beta_1 = 0$, then the Sylvester's criterion $\begin{vmatrix} \alpha & x_1 \\ x_1^* & \beta_1 \end{vmatrix} = \alpha\beta_1 - |x_1|^2 \geq 0$ forces $x_1 = 0$. In that situation the term $|x_1|^2/\beta_1$ is interpreted as 0; the inequality then reduces to the same statement with that term omitted, and it still holds. Thus, if the sum of the two-tangles vanishes, it provides a sufficient condition for the state to be separable, indicating that its entanglement has been completely lost. The resulting mixed separable state can be expressed in the form (15).

Next, the proof is extended to the $n$-qubit case. In general, the density matrix of an $n$-qubit state supported on the $W$ subspace can be written as follows:

$$\rho_n = \begin{pmatrix} \mathrm{A} & X_1 & \cdots & X_n \\ X_1^* & \mathrm{B}_{1,1} & \cdots & \mathrm{B}_{2,n} \\ \vdots & \vdots & \ddots & \vdots \\ X_n^* & \mathrm{B}_{n,2} & \cdots & \mathrm{B}_{n,n} \end{pmatrix}, \quad (26)$$

where $\mathrm{B}_{i,j} = \mathrm{B}_{j,i}^*$, and $\mathrm{A} + \sum_i \mathrm{B}_{i,i} = 1$. Also, the density matrix above is expressed in the basis given in Eq. (7). By performing $n-2$ partial traces, the reduced density matrices of subsystems $s$ and $r$ are obtained as follows:

$$\rho_n^{s,r} = \begin{pmatrix} \mathrm{A} + \sum_{m \neq r,s} \mathrm{B}_{m,m} & X_r & X_s & 0 \\ X_r^* & \mathrm{B}_{r,r} & \mathrm{B}_{r,s} & 0 \\ X_s^* & \mathrm{B}_{s,r} & \mathrm{B}_{s,s} & 0 \\ 0 & 0 & 0 & 0 \end{pmatrix}. \quad (27)$$

Using Eqs. (1) and (27), the sum of two-tangles is derived as follows:

$$\mathrm{Total}\left(C_{s,r}^2\right) = 4 \sum_{\substack{r,s=1 \\ r \neq s}}^{n} |\mathrm{B}_{s,r}|^2. \quad (28)$$

Therefore, if the sum of two-tangles vanishes, all the coherences $\mathrm{B}_{s,r}$ also vanish. Consequently, the density matrix $\rho_n$ transforms into the following form:

$$\rho'_n = \begin{pmatrix} \mathrm{A} & X_1 & \cdots & X_n \\ X_1^* & \mathrm{B}_{1,1} & \cdots & 0 \\ \vdots & \vdots & \ddots & \vdots \\ X_n^* & 0 & \cdots & \mathrm{B}_{n,n} \end{pmatrix}. \quad (29)$$

Now, it must be shown that the state $\rho'_n$ is a separable mixed state. To do so, an expansion of this state in the form of Eq. (15) is sought. For this purpose, the following states are considered:

$$|\psi_0\rangle = |0\rangle^{\otimes n}, \quad (30)$$
$$|\psi_i\rangle = \sqrt{1-|\mu_i|^2}\,|0\rangle^{\otimes n} + \mu_i e_i \quad (i \neq 0),$$

where $e_i$ is defined as follows:

$$e_i \in \left\{ |0\rangle^{\otimes(n-1)}|1\rangle, |0\rangle^{\otimes(n-2)}|10\rangle, \cdots, |1\rangle|0\rangle^{\otimes(n-1)} \right\}. \quad (31)$$

For example $e_1 = |0\rangle^{\otimes(n-1)}|1\rangle$. Using Eqs. (15) and (30) and comparing the result with density matrix in Eq. (29), the same relations as (17) to (20) are derived again.

$$p_0+\sum_{i=1}^{n} p_i\left|\upsilon_i\right|^2=\mathrm{A}, \tag{32}$$

$$p_i\left|\mu_i\right|^2=\mathrm{B}_{i,i}\quad(i\neq 0), \tag{33}$$

$$p_i\mu_i\sqrt{1-\left|\mu_i\right|^2}=X_i\quad(i\neq 0), \tag{34}$$

$$p_i=\mathrm{B}_{i,i}+\frac{\left|X_i\right|^2}{\mathrm{B}_{i,i}}\quad(i\neq 0). \tag{35}$$

Therefore, the second relation (30) can be rewritten as follows:

$$\left|\psi_i\right\rangle=\frac{1}{\sqrt{p_i}}\left(\frac{X_i}{\sqrt{\mathrm{B}_{i,i}}}\left|0\right\rangle^{\otimes n}+\sqrt{\mathrm{B}_{i,i}}e_i\right). \tag{36}$$

In the above relation $i$ is a positive integer. Thut, the probability coefficient $p_0$ can be calculated as follows:

$$p_0=\mathrm{A}-\sum_{i=1}^{n}\frac{\left|X_i\right|^2}{\mathrm{B}_{i,i}}. \tag{37}$$

Next, it must be shown that the following inequality always holds to ensure that the probability coefficient $p_0$ remains nonnegative.

$$\mathrm{A}\geq\sum_{i=1}^{n}\frac{\left|X_i\right|^2}{\mathrm{B}_{i,i}}. \tag{38}$$

Given that the density matrix $\rho_n'$ is positive semidefinite, meaning its determinant must be nonnegative, the following relation must hold:

$$Det(\rho_n')=\mathrm{A}\prod_{i=1}^{n}\mathrm{B}_{i,i}+\frac{\left|X_1\right|^2}{\mathrm{B}_{1,1}}\prod_{i=1}^{n}\mathrm{B}_{i,i} \ldots+\frac{\left|X_n\right|^2}{\mathrm{B}_{n,n}}\prod_{i=1}^{n}\mathrm{B}_{i,i}\geq 0. \tag{39}$$

Since the elements $\mathrm{B}_{i,i}$ are nonnegative and real, and assuming they are non-zero, we divide the above inequality by $\prod_{i=1}^{n}\mathrm{B}_{i,i}$ to obtain inequality (38). Again, the Sylvester's criterion guarantees that even if some of the elements $\mathrm{B}_{i,i}$ are zero, the above inequality still holds.

It should be noted that Theorem 1 is not restricted to the two-tangle. Even if the sum of pairwise negativities vanishes, the state becomes fully separable. Demonstrating this result using Theorem 1 is straightforward. To show that, let's denote each bipartition of the state by $\rho_n^{s,r}$. Because for two qubits, the PPT criterion is both necessary and sufficient for separability [18], if the negativity of $\rho_n^{s,r}$ vanishes, then that bipartition is separable. Consequently, its two-tangle also vanishes. Therefore, if the negativities of all bipartitions are zero, the sum of two-tangles is also zero. Hence, according to Theorem 1, the system is completely separable.

## IV. The sum of two-tangles as an entanglement measure for pure states supported in the *W* subspace

A valid measure of entanglement must satisfy the following conditions [19]:

a) Zero for separable states.
b) Local unitary invariance: The measure stays unchanged under any local unitary transformations.
c) Monotonicity: The measure does not increase when subjected to local operations and classical communication (LOCC).

The sum of two-tangles is then defined as a quantity and it is shown to satisfy the three conditions above. This thereby constitutes a valid entanglement measure for any pure state supported in the *W* subspace.

$$\Sigma\tau_2\left(\left|\psi\right\rangle\right)=Z\sum_{i<j}\tau_{i,j}. \tag{40}$$

In the above relation, $\Sigma\tau_2$ denotes the sum of two-tangles and $Z$ is a normalization constant that rescales the quantity to the interval $[0,1]$. The maximum value of the sum of two-tangles for pure states supported in the $W$ subspace is $2(n-1)/n$ (see Appendix A), with equality attained by the $n$-qubit $W$ state [20]. Therefore, we choose the normalization constant as

$$Z=\frac{n}{2(n-1)}, \tag{41}$$

such that the resulting measure is normalized to unity for the $W$ state.

Based on Theorem 1, the sum of two-tangles satisfies condition (a). Now it is shown that this quantity also satisfies conditions (b) and (c). The proof relies on the fact that each individual two-tangle already fulfills these conditions on its own. In an $n$-qubit state, the number of bipartitions is $n(n-1)/2$. Now, let $\tau_1(\rho)$, $\tau_2(\rho)$, ..., $\tau_{n(n-1)/2}(\rho)$ be all two-tangles on state $\rho$. Since, each two-tangle is invariant under local unitaries, i.e. for any local unitary $U=U_A\otimes U_B$, we have: $\tau_i(U\rho U^{\dagger})=\tau_i(\rho)$, then

$$\begin{aligned}\Sigma\tau_2\left(U\rho U^{\dagger}\right)&=\tau_1\left(U\rho U^{\dagger}\right)+\cdots\\+\tau_{n(n-1)/2}\left(U\rho U^{\dagger}\right)&=\tau_1(\rho)+\cdots\\+\tau_{n(n-1)/2}(\rho)&=\Sigma\tau_2(\rho).\end{aligned} \tag{42}$$

So $\Sigma\tau_2$ is invariant under local unitaries. For the proof of condition (c), see Appendix B.

Fig. 1 illustrates the sum of two-tangles of state (6). It is observed that for $\alpha=\beta=1/\sqrt{3}$, corresponding to the $W$ state, it reaches its maximum value. Considering the normalization constant $Z=3/4$, the entanglement of $W$ state is 1. For the special cases $k_1=0$ or $k_2=0$, the entanglement is non-zero. Because in this cases, state (6) becomes a biseparable state in which a pair of qubits remains entangled. For example, for $k_1=0$, state (6) transforms into $|SB\rangle\otimes|0\rangle$ where $|SB\rangle\equiv k_2|01\rangle+k_3|10\rangle$ is a semi-Bell state with $|k_1|^2+|k_2|^2=1$. In this special case, the maximum entanglement of the biseparable state is $3/4$.

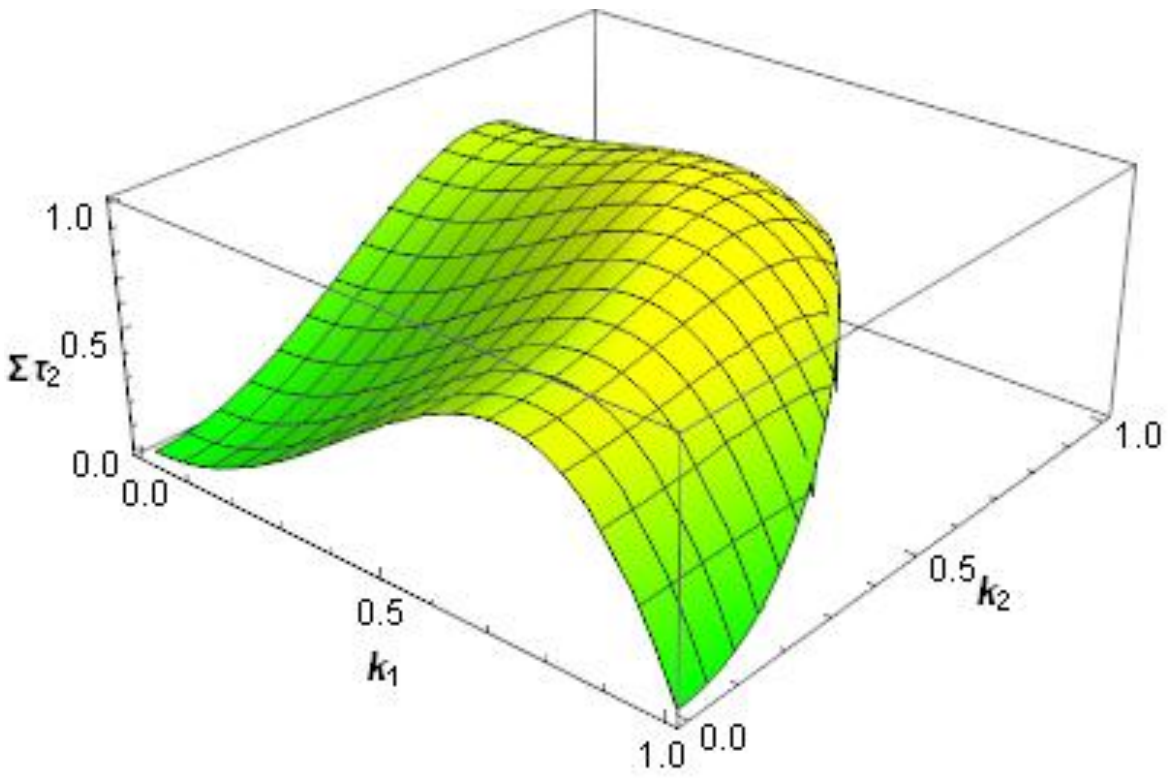

**FIG. 1:** The sum of two-tangles of state (6) assuming $Z=3/4$.

It is useful to evaluate the entanglement of the $n$-qubit $W$ state by computing both the sum of two-tangles and the π-tangle and then comparing the two results. Because the entanglement of the $W$ state depends on the number of qubits [20], it is natural to ask whether, as $n$ increases, the sum of two-tangles and the π-tangle continue to provide accurate measures of the total entanglement. Using Eq. (27) the density matrix of each bipartition $s$ and $r$ corresponding to the $n$-qubit $W$ state is obtained as follows:

$$\rho_{W_n}^{s,r}=\frac{1}{n}\begin{pmatrix}n-2&0&0&0\\0&1&1&0\\0&1&1&0\\0&0&0&0\end{pmatrix}. \tag{43}$$

Now, using Eqs. (1), (40) and (43), the sum of two-tangles is found to be:

$$\Sigma\tau_2\left(\rho_{W_n}\right)=\frac{2(n-1)}{Z\,n}. \tag{44}$$

Using Eq. (41), we obtain:

$$\Sigma\tau_2\left(\rho_{W_n}\right)=1. \tag{45}$$

Therefore, the sum of two-tangles for the $n$-qubit $W$ state is independent of the number of qubits.

On the other hand, the π-tangle for the $n$-qubit $W$ state is derived as follows [8]:

$$\pi_{W_n}=\left(\frac{2\sqrt{n-1}}{n}\right)^2 -(n-1)\left(\frac{\sqrt{(n-2)^2+4}-n+2}{n}\right)^2. \tag{46}$$

Fig. 2 shows the variation of the the π-tangle and the sum of two-tangles versus the number of qubits. It is observed that, as $n$ increases, the π-tangle tends to zero, while the sum of two-tangles remains equal to one for all $n$. This means that the π-tangle cannot capture the entanglement of the $W$ state for large $n$. However, this problem can be resolved by summing over all the π-tangles. So it can be stated that for a pure $n$-qubit state $|\psi\rangle$, the sum of π-tangles can be defined as follows:

$$\Sigma\pi\left(|\psi\rangle\right)=\frac{1}{4}\sum_i \pi_i. \tag{47}$$

Here, the symbol $\Sigma\pi$ is chosen to represent this measure. Moreover, the normalization constant is $1/4$. This choice is justified by the fact that the sum of π-tangles is universally bounded from above by 4 for all pure states supported in the $W$ subspace, as shown in Appendix C. Therefore, the normalized quantity satisfies $0\le\Sigma\pi\left(|\psi\rangle\right)\le 1$ for every $n$.

The proof that the sum of π-tangles is an entanglement measure is analogous to the proof for the sum of two-tangles. In contrast to the sum of two-tangles, which is applicable only to states in the $W$ subspace, the sum of π-tangles can be used for a wide range of $n$-qubit systems.

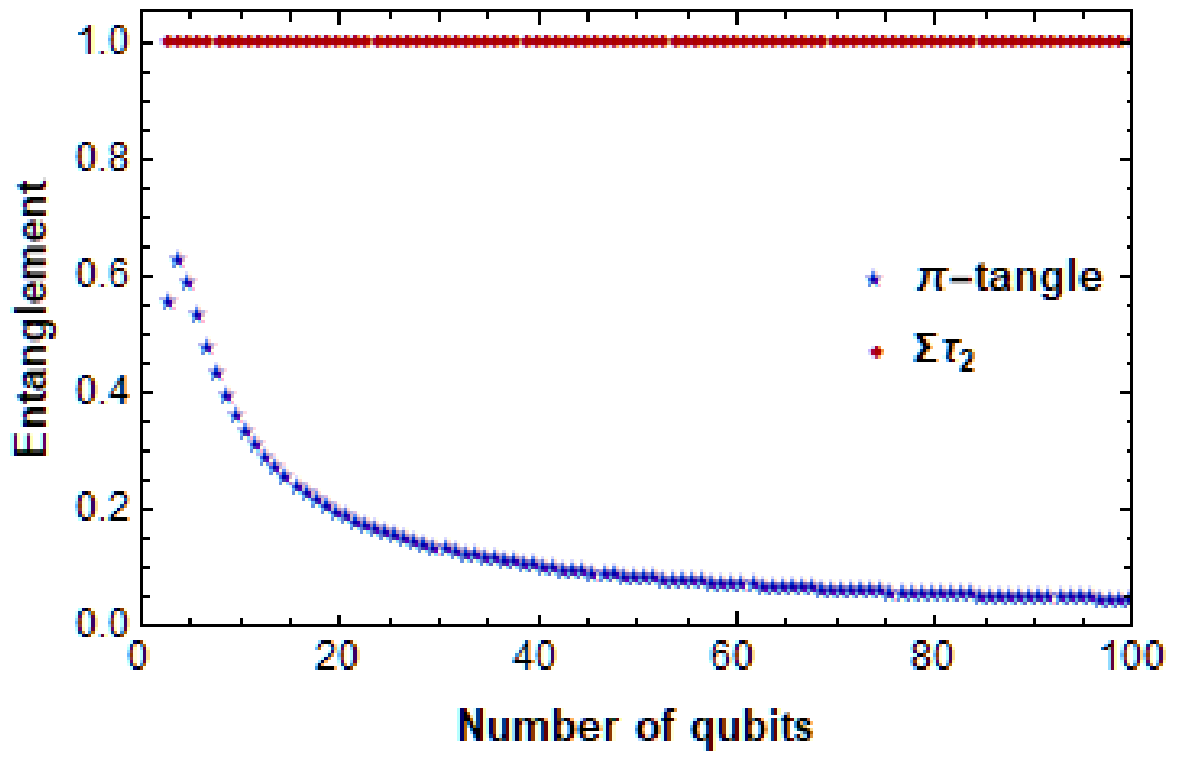

**FIG. 2:** The sum of two-tangles and the π-tangle of the $n$-qubit $W$ state versus the number of qubirts.

Using Eqs. (46) and (47), the sum of π-tangles for the $n$-qubit $W$ state is obtained as follows:

$$\Sigma\pi\left(|W\rangle\right)=\frac{n-1}{4n}\times\left\{4-\left(\sqrt{(n-2)^2+4}-n+2\right)^2\right\}. \tag{48}$$

Since $\underset{n\to\infty}{Limit}\,\Sigma\pi\left(\rho_W\right)=1$,the sum of the two-tangles does not vanish in the large-$n$ limit and can therefore quantify the entanglement of the $W$ state even as $n\to\infty$. A very important point to note is that the entanglement of some quantum states, such as the GHZ state, does not depend on the number of qubits. In particular, for families of states whose probability coefficients are independent of the number of qubits, the amount of entanglement remains unchanged as the number of qubits increases. In this case, the sum of π-tangles is not an appropriate measure. In fact, the use of this measure is important for states in which the π-tangle tends to zero as number of qubits increases.

The discussion above leads us to the definition of a new condition for entanglement measures. This condition can be stated as follows: Assuming an entanglement measure $T$ and an entangled $n$-qubit density matrix $\rho_n$, the following relation should be satisfied:

$$\underset{n\to\infty}{Limit}\, T\left(\rho_n\right) \neq 0. \qquad (49)$$

This condition stated that if the number of qubits increases, the entanglement measure $T$ should not tends to zero. It should be noted that, unlike the three necessary conditions for an entanglement measure, the above condition is not strictly necessary; rather, it serves to guide the definition of a well-behaved entanglement measure for $n$-qubit systems. Similarly, other non-essential conditions, such as convexity and additivity, have been proposed in the past [19]. Although not required, these criteria help construct entanglement measures with useful mathematical and operational properties.

## V. Conclusion

In this work, a separability criterion was established for any pure or mixed state supported in the $W$ subspace. Specifically, it was shown that if all pairwise entanglements vanish, then the $n$-qubit state is separable. In other words, the state can be written in the form of the ensemble decomposition given in Eq. (15). One application of this theorem is to determine the separability of a state obtained from the $W$ state through a transformation that preserves the basis of the subspace.

Based on Theorem 1, it was shown that the sum of two-tangles satisfies the necessary conditions of an entanglement measure for pure states supported in the $W$ subspace. The importance of defining such a measure becomes even more apparent when we consider that the three-tangle, given in Eq. (2), is not an entanglement measure for the $W$-class states since its value vanishes for them [10]. A key advantage of the sum of two-tangles is its computational simplicity, making it particularly useful for analytical calculations.

It was also demonstrated that for the $n$-qubit $W$ state, as the number of qubits increases, the π-tangle tends to zero, while the sum of two-tangles remains equal to one for all $n$. Consequently, the π-tangle fails to quantify the entanglement of the $n$-qubit $W$ state in the large-$n$ limit. To overcome this limitation, the sum of π-tangles is introduced as an entanglement measure, which effectively resolves the issue. Moreover, this observation led us to introduce a new condition for an arbitrary entanglement measure $T$, specified by Eq. (49). Although, in general, this condition is not necessary, it can contribute to the definition of a well-behaved entanglement measure. The reason the π-tangle cannot quantify the entanglement of the $n$-qubit $W$ state for large $n$ is that it does not satisfy this condition. Both the sum of two-tangles and the sum of π-tangles satisfy the condition (49), allowing them to effectively quantify the entanglement of the $n$-qubit $W$ state in the large-$n$ limit.

## APPENDIX A

We need to show that the maximum value of the sum of two-tangles among pure states in the $W$ subspace is $2(n-1)/n$. A general pure state in the $W$ subspace is

$$|\psi\rangle = a_0 |0\rangle^{\otimes n} + \sum_{i=1}^{n} a_i |i\rangle, \qquad (A1)$$

where $|i\rangle$ denotes the state with the only excitation at qubit $i$, and

$$|a_0|^2 + \sum_{i=1}^{n} |a_i|^2 = 1. \qquad (A2)$$

For arbitrary qubits $i$ and $j$, the reduced two-qubit state is

$$\rho_{i,j} = \begin{pmatrix} 1-|a_i|^2-|a_j|^2 & a_0 a_j^* & a_0 a_i^* & 0 \\ a_j a_0^* & |a_j|^2 & a_j a_0^* & 0 \\ a_i a_0^* & a_j a_0^* & |a_i|^2 & 0 \\ 0 & 0 & 0 & 0 \end{pmatrix}. \qquad (A3)$$

Using Eq. (1), the two-tangle is obtained as follows:

$$\tau_2\left(|\psi\rangle\right)=4|a_i|^2|a_j|^2=4p_ip_j, \tag{A4}$$

where $p_i=|a_i|^2$ and $p_j=|a_j|^2$. On the other hand, using Eq. (A2), we obrain

$$\sum_{i=1}^{n}p_i=1-|a_0|^2\le 1. \tag{A5}$$

Now, using

$$\left(\sum_{i=1}^{n}p_i\right)^2=\sum_i p_i^2+2\sum_{i<j}p_ip_j, \tag{A6}$$

and Eq. (A4), we find

$$\begin{aligned}\Sigma\tau_2\left(|\psi\rangle\right)&\equiv 4\sum_{i<j}^{n}p_ip_j\\ &=2\left\{\left(\sum_{i=1}^{n}p_i\right)^2-\sum_{i=1}^{n}p_i^2\right\}.\end{aligned} \tag{A7}$$

Now, we define

$$P=\sum_{i=1}^{n}p_i=1-|a_0|^2. \tag{A8}$$

Then, by the Cauchy–Schwarz inequality

$$\left(\sum_{i=1}^{n}p_i\right)^2\le n\sum_{i=1}^{n}p_i^2, \tag{A9}$$

it follows that

$$\frac{P^2}{n}\le\sum_{i=1}^{n}p_i^2. \tag{A10}$$

Using Eqs. (A7)-(A10), we find

$$\Sigma\tau_2\left(|\psi\rangle\right)\le 2\left(P^2-\frac{P^2}{n}\right)=\frac{2(n-1)}{n}P^2. \tag{A11}$$

Since $P\le 1$, then

$$\Sigma\tau_2\left(|\psi\rangle\right)\le\frac{2(n-1)}{n}. \tag{A12}$$

## APPENDIX B

For pure states supported in the $W$ subspace, the one-tangle $\tau_i$ satisfies [21]

$$\tau_i=\sum_{j\ne i}\tau_{i,j}. \tag{B1}$$

In the above equation, $i$ is fixed, and the summation is taken over $j$. Also, $\tau_{i,j}$ represents the two-tangle between qubits $i$ and $j$. This equality is a characteristic property of $W$-class states, for which the 3-tangle vanishes [10, 21]. Consequently,

$$\sum_i\tau_i=2\sum_{i<j}\tau_{i,j}. \tag{B2}$$

Therefore, if the sum of one-tangles is non-increasing on average under LOCC, then the sum of two-tangles is also non-increasing on average. Indeed,

$$\sum_\mu p_\mu\sum_i\tau_i\left(\psi_\mu\right)\le\sum_i\tau_i\left(\psi\right), \tag{B3}$$

Implies

$$\sum_\mu p_\mu\sum_{i<j}\tau_{i,j}\left(\psi_\mu\right)\le\sum_{i<j}\tau_{i,j}\left(\psi\right). \tag{B4}$$

Here, $\mu$ labels the possible outcomes, $p_\mu$ is the probability of obtaining outcome $\mu$, and $\psi_\mu$ denotes the corresponding post-measurement state.

It therefore remains to establish the LOCC monotonicity of each one-tangle. For this purpose, consider the bipartition $i|\text{rest}$. The one-tangle is given by [21]

$$\tau_i=4\det\left(\rho_i\right), \tag{B5}$$

where $\rho_i$ is the reduced density matrix of qubit $i$. It has been shown that, for a bipartite pure state $|\psi\rangle_{AB}$, an entanglement monotone can be constructed from a function $f\left(\rho_A\right)$ of its reduced density matrix $Tr_B\left(|\psi\rangle\langle\psi|\right)$, provided that $f$ is unitarily invariant and

concave [22-23]. For the one-tangle, we define

$$f(\rho_i) = 4\det(\rho_i). \quad \text{(B6)}$$

We need to show that this function satisfies the required properties.

**Unitary invariance**: For any unitary operator *U*, we have:

$$f(U\rho_i U^\dagger) = 4\det(U\rho_i U^\dagger) = 4\det(\rho_i) = f(\rho_i). \quad \text{(B7)}$$

Then, $f(\rho_i)$ remains unchanged under unitary transformations.

**Concavity.** A function $f$ is concave if [22]

$$f(\lambda\rho_1 + (1-\lambda)\rho_2) \geq \lambda f(\rho_1) + (1-\lambda) f(\rho_2), \quad \text{(B8)}$$

where $0 \leq \lambda \leq 1$. Using the below relation [24]

$$f(\rho_i) = 4\det(\rho_i) = 2\left(1 - Tr(\rho_i^2)\right), \quad \text{(B9)}$$

one obtains

$$\begin{aligned} &f(\lambda\rho_1 + (1-\lambda)\rho_2) \\ &-\lambda f(\rho_1) - (1-\lambda) f(\rho_2) \\ &= 2\lambda(1-\lambda)\mathrm{Tr}\left[(\rho_1 - \rho_2)^2\right] \geq 0. \end{aligned} \quad \text{(B10)}$$

Thus, $f(\rho_i)$ is concave. Therefore,

$$\sum_\mu p_\mu \tau_i(\psi_\mu) \leq \tau_i(\psi). \quad \text{(B11)}$$

Summing the above inequality over *i* gives inequality (B3). Then, using Eq. (B2), inequality (B4) follows. Since any finite-round LOCC protocol can be decomposed into a sequence of local operations, and the sum of pairwise two-tangles is non-increasing on average under each individual local operation, it is also non-increasing on average under the complete LOCC protocol [25].

## APPENDIX C

For the general pure state in the *W* subspace given in Eq. (A1), we define

$$\begin{aligned} p_0 &= |a_0|^2, \\ p_i &= |a_i|^2. \end{aligned} \quad \text{(C1)}$$

The squared negativity $N^2_{i|\mathrm{rest}}$ is equivalent to the one-tangle when the latter is defined in terms of negativity. In particular for pure state we have [12]:

$$N^2_{i|\mathrm{rest}} = 4\det(\rho_i), \quad \text{(C2)}$$

where $\rho_i$ is the single-qubit reduced density matrix. Moreover, $\rho_i$ can be obtained by tracing out qubit *j* from the density matrix in Eq. (A3).

$$\rho_i = \begin{pmatrix} 1-p_i & a_0 a_i^* \\ a_i a_0^* & p_i \end{pmatrix}. \quad \text{(C3)}$$

Using Eqs. (C1)-(C3), we obtain

$$N^2_{i|\mathrm{rest}} = 4p_i(1 - p_0 - p_i). \quad \text{(C4)}$$

Hence, the sum of one-tangles is

$$\sum_{i=1}^n N^2_{i|\mathrm{rest}} = 4\sum_{i=1}^n p_i(1 - p_0 - p_i). \quad \text{(C5)}$$

Using

$$\sum_{i=1}^n p_i = 1 - p_0, \quad \text{(C6)}$$

we obtain

$$\begin{aligned} \sum_{i=1}^n N^2_{i|\mathrm{rest}} &= 4\left\{(1-p_0)\sum_{i=1}^n p_i - \sum_{i=1}^n p_i^2\right\} \\ &= 4\left\{(1-p_0)^2 - \sum_{i=1}^n p_i^2\right\}. \end{aligned} \quad \text{(C7)}$$

Since $p_i \geq 0$, we have:

$$\sum_{i=1}^{n} p_i^2 \geq 0. \tag{C8}$$

Therefore

$$\sum_{i=1}^{n} N_{i|\text{rest}}^2 \leq 4\left(1-p_0\right)^2 \leq 4. \tag{C9}$$

Thus,

$$\sum_{i=1}^{n} N_{i|\text{rest}}^2 \leq 4. \tag{C10}$$

Baed on the monogamy of entanglement, we have [26]:

$$\pi_i = N_{i|\text{rest}}^2 - \sum_{i \neq j} N_{i,j}^2. \tag{C11}$$

Summing over *i*, we obtain

$$\Sigma\pi \equiv \sum_{i=1}^{n} \pi_i = \sum_{i=1}^{n} N_{i|\text{rest}}^2 - 2\sum_{i<j} N_{i,j}^2. \tag{C12}$$

Since every two-tangle satisfies $N_{i,j} \geq 0$,

$$\Sigma\pi \leq \sum_{i=1}^{n} N_{i|\text{rest}}^2 \leq 4. \tag{C13}$$

Therefore,

$$0 \leq \Sigma\pi \leq 4. \tag{C14}$$

In other words

$$0 \leq \frac{1}{4}\Sigma\pi \leq 1. \tag{C15}$$

## DATA AVAILABILITY STATEMENT

The data that support the findings of this study are available from the corresponding author, [R. Hamzehofi], upon reasonable request.

## ACKNOWLEDGMENTS

This research has received no external funding.

## References

[1] Y. Yeo and W. K. Chua, Teleportation and dense coding with genuine multipartite entanglement. Phys. Rev. Lett. **96**, 060502 (2006). https://doi.org/10.1103/PhysRevLett.96.060502

[2] M. Choi, E. Bae, and S. Lee, Genuine multipartite entanglement measures based on multi-party teleportation capability. Sci. Rep. **13**, 15013 (2023). https://doi.org/10.1038/s41598-023-42052-x

[3] W.-T. Kao, C.-Y. Huang, T.-J. Tsai, S.-H. Chen, S.-Y. Sun, Y.-C. Li, T.-L. Liao, C.-S. Chuu, H. Lu, and C.-M. Li, Scalable determination of multipartite entanglement in quantum networks. npj Quantum Inf. **10**, 73 (2024). https://doi.org/10.1038/s41534-024-00867-0

[4] A. Chakraborty, R. K. Patra, K. Agarwal, S. Sen, P. Ghosal, S. G. Naik, and M. Banik, Scalable and noise-robust communication advantage of multipartite quantum entanglement. Phys. Rev. A **111**, 032617 (2025). https://doi.org/10.1103/PhysRevA.111.032617

[5] L. Oleynik, J. ur Rehman, S. Koudia, and S. Chatzinotas, Entanglement distribution in lossy quantum networks. Sci. Rep. **15**, 29778 (2025). https://doi.org/10.1038/s41598-025-14226-2

[6] K. Berrada and S. Bougouffa, Quantum W-type entanglement in photonic systems with environmental decoherence. Symmetry **17**, 1147 (2025). https://doi.org/10.3390/sym17071147

[7] H. Haeffner, W. Haensel, C. F. Roos, J. Benhelm, D. Chek-al-kar, M. Chwalla, T. Koerber, U. D. Rapol, M. Riebe, P. O.

Schmidt, C. Becher, O. Gühne, W. Dür, R. Blatt Berrada, Scalable multi-particle entanglement of trapped ions,Nature 438, 643-646 (2005). https://doi.org/10.1038/nature04279

[8] R. Hamzehofi, M. Ashrafpour, and D. Afshar, Genuine entanglement and quantum coherence of a multipartite W state in non-inertial frames. Eur. Phys. J. Plus **140**, 986 (2025). https://doi.org/10.1140/epjp/s13360-025-06942-5

[9] W. K. Wootters, Entanglement of formation of an arbitrary state of two qubits. Phys. Rev. Lett. **80**, 2245 (1998). https://doi.org/10.1103/PhysRevLett.80.2245

[10] V. Coffman, J. Kundu, and W. K. Wootters, Distributed entanglement. Phys. Rev. A **61**, 052306 (2000). https://doi.org/10.1103/PhysRevA.61.052306

[11] Y.-C. Ou and H. Fan, Monogamy inequality in terms of negativity for three-qubit states. Phys. Rev. A **75**, 062308 (2007). https://doi.org/10.1103/PhysRevA.75.062308

[12] G. Vidal and R. F. Werner, Computable measure of entanglement. Phys. Rev. A **65**, 032314 (2002). https://doi.org/10.1103/PhysRevA.65.032314

[13] W. Dür, G. Vidal, and J. I. Cirac, Three qubits can be entangled in two inequivalent ways. Phys. Rev. A **62**, 062314 (2000). https://doi.org/10.1103/PhysRevA.62.062314

[14] D. Bao, M. Liu, Y. Ou, Q. Xu, Q. Li, and X. Tan, Eigenvalue-based quantum state verification of three-qubit W class states. Physica A **639**, 129681 (2024). https://doi.org/10.1016/j.physa.2024.129681

[15] D. McMahon, Quantum Computing Explained, Wiley-Interscience (Hoboken), (2008).

[16] M. A. Nielsen and I. L. Chuang, Quantum Computation and Quantum Information, 10th Anniversary Ed., Cambridge Univ. Press (Cambridge), (2010).

[17] G. T. Gilbert, Positive definite matrices and Sylvester's criterion. Am. Math. Mon. **98**, 44 (1991). https://doi.org/10.2307/2324036

[18] A. Peres, Separability criterion for density matrices. Phys. Rev. Lett. **77**, 1413 (1996). https://doi.org/10.1103/PhysRevLett.77.1413

[19] M. B. Plenio and S. Virmani, An introduction to entanglement measures. Quantum Inf. Comput. **7**, 1 (2007). https://doi.org/10.26421/QIC7.1-2-1

[20] R. Hamzehofi, Entanglement measures for multipartite quantum systems: limitations and new approaches, Eur. Phys. J. D **80**, 67 (2026) https://doi.org/10.1140/epjd/s10053-026-01182-3

[21] P. J. Geetha, K. O. Yashodamma and Sudha, Monogamous nature of symmetric N-qubit states of the W class: Concurrence and negativity tangle, Chin. Phys. B **24**, 11 (2015). doi:10.1088/1674-1056/24/11/110302

[22] G. Vidal, Entanglement monotones, J. Mod. Opt. **47**, 355 (2000). https://doi.org/10.1080/09500340008244048

[23] R. Horodecki, P. Horodecki, M. Horodecki, and K. Horodecki, Quantum entanglement, Rev. Mod. Phys. **81**, 865 (2009). https://doi.org/10.1103/RevModPhys.81.865

[24] V. S. Bhaskara, P. K. Panigrahi, Generalized concurrence measure for faithful quantification of multiparticle pure state entanglement using Lagrange's identity and wedge product, Quantum Inf. Process. **16**, 118 (2017). https://doi.org/10.1007/s11128-017-1568-0

[25] O. Oreshkov, T. A. Brun, Infinitesimal local operations and differential conditions for entanglement monotones, Phys. Rev. A **76**, 059905 (2007). https://doi.org/10.1103/PhysRevA.73.042314

[26] L.-M. Gao, F.-L. Yan, T. Gao, Monogamy of logarithmic negativity and logarithmic convex-roof extended negativity, Results Phys. **31**, 104983 (2021). https://doi.org/10.1016/j.rinp.2021.104983